\documentclass[
    ,final            
  ]
  {aipproc}

\layoutstyle{8x11single}

\begin{document}

\title{Interaction matrix element fluctuations in quantum dots}



\author{L. Kaplan}{address={Department of Physics, Tulane University, New Orleans, LA 70118, USA}}

\author{Y. Alhassid}{address={Center for Theoretical Physics, Sloane Physics Laboratory, Yale
University, New Haven, CT 06520, USA}}

\begin{abstract}
 In the Coulomb blockade regime of a ballistic quantum dot, the distribution of conductance peak spacings is
 well known to be incorrectly predicted by a single-particle picture; instead, matrix element fluctuations
 of the residual electronic interaction need to be taken into account. In the normalized random-wave model,
 valid in the semiclassical limit where the number of electrons in the
 dot becomes large, we obtain analytic expressions for the fluctuations of two-body and one-body matrix elements.
 However, these fluctuations may be too small to explain low-temperature experimental data.
 We have examined matrix element fluctuations in realistic chaotic geometries, and shown that at energies of
 experimental interest these fluctuations generically exceed by a factor of about 3-4 the predictions of the
 random wave model. Even larger fluctuations occur in geometries with a mixed chaotic-regular phase space.
 These results may allow for much better agreement between the Hartree-Fock picture and experiment. Among
 other findings, we show that the distribution of interaction matrix elements is strongly non-Gaussian in
 the parameter range of experimental interest, even in the random wave model. We also find that the
 enhanced fluctuations in realistic geometries cannot be computed using a leading-order semiclassical
 approach, but may be understood in terms of short-time dynamics.
\end{abstract}

\maketitle

\section{Introduction}

 There has been much interest in the mesoscopic properties of quantum dots
whose single-particle dynamics are chaotic \cite{alhassid00}.  The generic
fluctuation properties of the single-particle spectrum and wave functions in
such dots are usually described by random matrix theory (RMT) \cite{guhr98}.
However, in almost-isolated dots, electron-electron interactions are important
and must also be taken into account.  The simplest model is the constant interaction
(CI) model, in which the interaction is taken to be the classical charging
energy.  Charging energy leads to Coulomb blockade peaks in the conductance
versus gate voltage.  Each peak occurs as the gate voltage is tuned to
compensate for the Coulomb repulsion and an additional electron tunnels into
the dot.  For a fixed number of electrons, the CI model is essentially a
single-particle model and RMT can be used to derive the statistical properties
of the conductance peak heights \cite{jalabert92}.  While the CI plus RMT model
has explained (at least qualitatively) \cite{jalabert92,alhassid96,alhassid98}
several observed features of the peak height fluctuations
\cite{folk96,chang96,folk01}, there have been significant discrepancies with
experimental data, in particular regarding the peak spacing statistics
\cite{sivan96,simmel97,patel98a,luscher01}.  Such discrepancies indicate the
importance of interactions beyond charging energy.

 A more systematic way of treating electron-electron interactions in chaotic
ballistic dots is to expand the interaction in a small parameter, the inverse
of the Thouless conductance $g_T \sim k L \sim \sqrt{N}$, where $k$ is the Fermi wave number,
$L$ is the typical linear size of the dot (i.e. $L=\sqrt{V}$, where $V$ is the area), and $N$ is the number of electrons in the dot.  In the limit of large Thouless
conductance (equivalently, in the semiclassical or many-electron limit), only a few interaction terms survive, constituting the interacting
part of the universal Hamiltonian \cite{kurland00,aleiner02}.  These universal
interaction terms include, in addition to charging energy, a constant exchange
interaction.  The inclusion of an exchange interaction has explained the
statistics of peak heights at low and moderate temperatures as well as the
suppression of the peak spacing fluctuations \cite{alhassid03,usaj03}.
However, at low temperatures, the peak spacing distribution remains bimodal
even when the exchange interaction is included, while none of the experimental
distributions are bimodal \cite{sivan96,patel98a,simmel97,luscher01}.

 For finite Thouless conductance, residual interactions beyond the universal
Hamiltonian must be taken into account. In a Hartree-Fock-Koopmans \cite{koopmans34}
approach (assuming the Hartree-Fock single-particle wave functions do not
change as electrons are added to the dot), the peak spacings can be expressed
directly in terms of certain (diagonal) two-body interaction matrix elements
\cite{alhassid02}.  Sufficiently large fluctuations of these interaction matrix
elements can explain the absence of bimodality in the peak spacing distribution
\cite{alhassid02,usaj02}.  In a diffusive
dot, the
variance of the matrix elements of the screened Coulomb interaction was shown
to behave as $\Delta^2/g_T^2$ to leading order in $1/g_T$~\cite{blanter97}, where the
single-electron mean level spacing $\Delta$ sets the energy scale.  However, dots studied in the
experiments are usually ballistic.

 An additional contribution to the peak spacing fluctuations originates in
surface charge effects \cite{blanter97}.  In a finite size system, screening
leads to the accumulation of charge on the surface of the dot.  The confining
one-body potential is then modified upon the addition of an electron to the
dot.

 In this paper we will summarize some of our recent results on fluctuations of the two-body interaction matrix
elements and of the surface charge one-body matrix elements in ballistic dots~\cite{kapalhinprep}.
We begin by defining the matrix elements of interest, and 
 note that their fluctuations can be expressed to leading order in $1/g_T$ in terms
 of spatial correlations within single-electron wave functions. 
  Berry's random wave conjecture~\cite{berry77} provides the first approximation
   for these correlations, which (in contrast with the situation for diffusive dots) is geometry
independent.  However, the spatial correlator of wave function
intensity obtained from the Berry conjecture is not consistent with the normalization requirement of the wave
functions \cite{gornyi02}.  We discuss the importance of normalization corrections to the
random wave correlator, and show how the variances of the two-body and one-body interaction matrix elements
may be computed  in a normalized random wave model. We also find that the distribution of interaction matrix
elements may be very far from Gaussian, even in a normalized random wave model where the wave functions are
very close to Gaussian.

An interesting quantity that we refrain from discussing here is the covariance
of interaction matrix elements, relevant for understanding spectral scrambling
when several electrons are added to the dot \cite{scrambling}. 
 
 We then proceed to study matrix element fluctuations in actual chaotic systems,
using a family of modified quarter-stadium billiards as an example.
We find strongly enhanced fluctuations in comparison with the normalized random wave
model.  Semiclassical corrections due to bounces from the dot's boundaries lead
to only a modest increase in the fluctuations, and do not correctly predict the
scaling with $kL$ in the experimentally relevant range.  Insight into the
underlying mechanism of fluctuation enhancement is obtained by examining a
family of quantum maps.  An important conclusion is that the expansion in $1/g_T$,
while asymptotically correct, can be problematic in quantifying matrix element
fluctuations in the regime relevant to experiments.  Finally, in the last section
we study systems beyond the chaotic regime, i.e., billiards
dominated by marginally-stable bouncing-ball modes as well as billiards with
mixed dynamics (partly regular and partly chaotic). 
  
\section{Interaction Matrix Elements}

The general two-body interaction matrix element for potential $v(\vec r',\vec r)$ is given
by
\begin{equation}
v_{\alpha\beta;\gamma\delta}\sim \int_V \int_V d\vec r \,d \vec r' \,\psi^\ast_\alpha(\vec r)\,\psi^\ast_\beta(\vec r')
\,v(\vec r',\vec r)\,\psi_\gamma(\vec r)\,\psi_\delta(\vec r')\,,
\end{equation}
where $\psi_\alpha$, $\psi_\beta$, $\ldots$ are single-electron orbital wave functions. In practice, screening
of the residual electron-electron interaction causes the range of the interaction to be much smaller than the size of the dot.
Modeling the residual interaction as a contact interaction, we obtain
$
v_{\alpha\beta;\gamma\delta}=\Delta\, V \int_V d\vec r \, \psi^\ast_\alpha(\vec r)\,\psi^\ast_\beta(\vec r)\,
\psi_\gamma(\vec r)\,\psi_\delta(\vec r)
$,
where the mean single-particle level spacing $\Delta$ sets the energy scale. We note that
for a contact interaction, exchange terms have precisely the same form and need not be considered separately.

Three distinct situations must be treated. First, the diagonal matrix elements 
\begin{equation}
v_{\alpha\beta}=v_{\alpha\beta;\alpha\beta}=\Delta V \int_V d\vec r \, |\psi_\alpha(\vec r)|^2\, |\psi_\beta(\vec r)|^2
\end{equation}
 ($\alpha \ne \beta$) appear when two electrons are found on the same two orbitals $\alpha$ and $\beta$ before and after the interaction. Secondly, we have the double-diagonal matrix elements $v_{\alpha\alpha}=v_{\alpha\alpha;\alpha\alpha}$, where two electrons are annihilated from one orbital and created on the same orbital (for a contact interaction,
 $v_{\alpha\alpha}$ is simply
 an inverse participation ratio in position space). Finally, we may consider the off-diagonal matrix elements $v_{\alpha\beta\gamma\delta}$ where the four orbitals are all distinct. We will focus mostly on the diagonal matrix elements $v_{\alpha\beta}$, but the other two situations are treated similarly.

The mean $\overline{v_{\alpha\beta}}$, averaged over all pairs of orbitals or over an ensemble, may always be absorbed into the mean
field part of the Hamiltonian. Thus, we are primarily interested not in the average, but in the fluctuations. To leading order in $g_T \sim kL$, the dominant
contribution to the variance arises from correlations between the intensities
of a {\it single} wave function at different points \cite{scrambling}:
\begin{equation}
\label{vabint}
\overline{\delta v_{\alpha\beta}^2} =
\Delta^2 V^2 \int_V \int_V d\vec r \, d\vec r' \,
C^2(\vec r,\vec r') +\cdots
\end{equation}
where 
\begin{equation}
C(\vec r,\vec r') 
= \overline{
|\psi(\vec r)|^2 \,
|\psi(\vec r')|^2 } -
\overline{|\psi(\vec r)|^2} \, \overline{|\psi(\vec r')|^2}
\,.
\label{cnorm1}
\end{equation}
Thus, the problem of two-body matrix elements statistics has been reduced to an apparently simpler problem of understanding the statistics of individual single-electron wave functions in a chaotic potential.
Terms we have omitted in Eq.~(\ref{vabint}) involve spatial correlators between $|\psi_\alpha(\vec r)|^2$ and $|\psi_\beta(\vec r')|^2$; such correlators are essential for computing the covariance $\overline{\delta v_{\alpha\beta}\delta v_{\alpha\gamma}}$, but are subleading in the calculation of the variance.

\section{Random Wave Approximation}\label{random-wave}

 For a two-dimensional billiard system, the random wave model implies that a
typical chaotic wave function may be written locally as a random superposition
of plane waves at fixed energy $\hbar^2k^2/2m$. Adopting the usual normalization
$\overline{|\psi(\vec r)|^2}={1/V}$, we obtain an amplitude correlator
$
\overline{\psi^\ast(\vec r) \psi(\vec r')}= {1 \over V} J_0(k|\vec r- \vec r'|)
$
and intensity correlator
\begin{equation}
C^{\rm rw}(\vec r,\vec r') = {1 \over V^2} {2 \over \beta} J_0^2(k|\vec r-
\vec r'|) \;, \label{corr}
\end{equation}
where $\beta=1$ or $2$ represents the presence or absence of time reversal symmetry, i.e.
the absence or presence of an external magnetic field, respectively.

The intensity correlator $C^{\rm rw}(\vec r,\vec r')$ is valid to leading order
in $|\vec r - \vec r'| /L$, but becomes problematic when applied to to all
$\vec r$, $\vec r'$ in the finite area $V$.  Indeed wave function
normalization requires the correlator to vanish on average,
\begin{equation}\label{normalization}
\int_V d \vec r \; C(\vec r,\vec r')=  0\,.
\end{equation}
However, $C^{\rm rw}(\vec r, \vec r')$ in (\ref{corr}) manifestly
does not satisfy the condition (\ref{normalization}).  The
reason for this failure is that in the random wave model, normalization is
satisfied only for the ensemble average, i.e., $\overline{\int_V d \vec r \, |\psi(\vec r)|^2
}= 1$, but not for each individual random superposition of plane waves $\psi(\vec r)$.

 This deficiency can be corrected by introducing the {\em normalized} random
wave model, in which each ``random'' wave function  is
normalized in area $V$, i.e.,
\begin{equation}
\psi^{\rm norm}(\vec r) = \psi(\vec r) /\int_V d \vec r \; |\psi(\vec r)|^2 \,.
\end{equation}
Assuming the deviation from exact normalization in area $V$ is small for the original
random waves $\psi(\vec r)$, we apply a perturbative
scheme~\cite{kapalhinprep} and obtain
\begin{equation}
\label{cnormexp}
C^{\rm norm}(\vec r,\vec r')= \tilde C (\vec r,\vec r')+ O\left ({ 1 \over
(kL)^{3/2}}\right) \,,
\end{equation}
where 
\begin{equation}
\tilde C(\vec r,\vec r') = C^{\rm rw}(\vec r,\vec r') - {1 \over V}
\int_V d\vec r_a \, C^{\rm rw}(\vec r,\vec r_a)  - {1 \over V}
\int_V d\vec r_a \, C^{\rm rw}(\vec r_a,\vec r')  + {1 \over V^2} \int_V 
\int_V d\vec r_a d\vec r_b \,C(^{\rm rw}\vec r_a,\vec r_b) \,.
\label{corrnorm}
\end{equation}
The leading-order normalized correlator $\tilde C(\vec r,\vec r')$ was previously derived
in Ref.~\cite{gornyi02} by adding a weak smooth disorder and using the
non-linear supersymmetric sigma model.
Higher-order contributions may be computed systematically; the full expression for the normalized two-point
correlator $C^{\rm norm}(\vec r, \vec r')$ involves all unnormalized $n$-point correlators
$\overline{(|\psi(\vec r_1)|^2-{1 \over V}) \cdots (|\psi(\vec r_n)|^2-{1 \over V})}$, starting with the unnormalized
three-point correlator that gives rise to the $O((kL)^{-3/2})$ correction in
Eq.~(\ref{cnormexp}). In practice the higher-order terms are small; the approximation $\tilde C$ already satisfies (\ref{normalization}) and can be proven to imply
normalization of individual wave functions~\cite{kapalhinprep}. However, $n$-point
correlators will be
important below, when we discuss the shape of matrix element distributions.

\subsection{Two-body matrix elements}\label{two-body}

Substituting the normalized random wave correlator $\tilde C$ into Eq.~(\ref{vabint}), we obtain
\begin{equation}
\overline{\delta v_{\alpha\beta}^2}  = \Delta^2 {3 \over \pi} \left({2 \over \beta} \right)^2 {\ln kL
+b_g\over (kL)^2} + O\left({\Delta^2 \over (kL)^3}\right) \,,
\label{bcoeff}
\end{equation}
where the leading $\ln kL/(kL)^2$ term may be obtained already from the unnormalized correlator $C^{\rm rw}$ and
depends only on the symmetry class, while
the shape-dependent coefficient $b_g$ may be easily evaluated by an integral involving Bessel functions over the area
$V$ of interest. 

\begin{figure}[ht] 
{
\includegraphics[height=.3\textheight,angle=270]{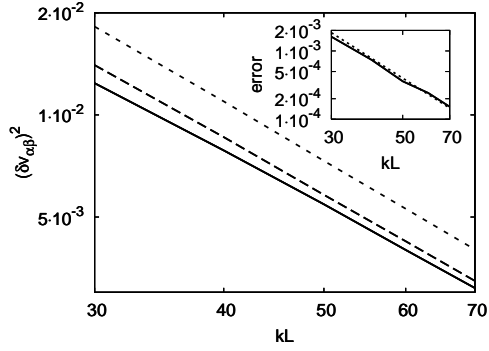}
}
\protect\caption{
The variance of the two-body matrix element $v_{\alpha\beta}$ versus $kL$ in
the real ($\beta=1$) random wave model: (a) the solid curve is the result of
exact numerical simulations; (b) the long-dashed line is the result of substituting the
normalized random wave correlator $\tilde C(\vec r,\vec r')$ into
Eq.~(\ref{vabint}); (c) the short-dashed line is the result of using the
unnormalized correlator $C^{\rm rw}(\vec r,\vec r')$.  In the inset, the solid line is
the difference between the full result (a) and the approximation (b); the
dashed line indicates that omitted terms scale as ${1 \over (kL)^3}$.  Here and
in all following figures, the single-particle level spacing $\Delta$, which sets the overall
energy scale, is set to unity.}
\label{figrandvab}
\end{figure}

In Fig.~\ref{figrandvab} we show the results for a disk geometry ($b_g=-0.1$) in the range
$30 \le kL \le 70$, corresponding roughly to the parameter range relevant for
experiments (i.e., a few hundred electrons in the dot). The discrepancy between the analytic result (\ref{bcoeff}) and exact numerics is indeed $O(1/(kL)^3)$, resulting in an accuracy of $\sim 10\%$ for the analytic formula. In practice, the dependence of $b_g$ on the geometry is weak, e.g. deforming the disk to an ellipse with an aspect ratio of $16$ while keeping the area fixed produces a change of only $\sim 5-6\%$ in the
$v_{\alpha\beta}$ variance for the experimental range of $kL$.
Thus, for all practical purposes, shape effects on the $v_{\alpha\beta}$
variance can be ignored (at least within the normalized random wave model).

Also, at this subleading order, we must in principle take into account the
wave number difference $\delta k=k_\alpha-k_\beta$. For $(k_\alpha+k_\beta)/2 \gg \delta k  \gg 1/L$ (i.e. for a wave number difference that is quantum mechanically large but classically small), this results merely in
a modification of the geometry-dependent coefficient in Eq.~(\ref{bcoeff}):
$b_g \to b_g - (1/3) \ln \delta k \,L$.  In practice, for
reasonable separations $\delta k \,L$, the
consequent reduction in the variance is at most $10\%$, and may be safely
ignored compared to the much larger dynamical effects to be discussed later.

 The variance of double-diagonal or off-diagonal matrix elements may be computed similarly.
To leading order, $O\left({\ln kL / (kL)^2}\right)$, all the results arise
from integrating the unnormalized correlator (\ref{corr}) and differ only by
geometry-independent combinatorial prefactors. However, normalization-related subtraction gives different results in the
three situations, and thus the subleading coefficient $b_g$ in (\ref{bcoeff}) must be replaced with
$b'_g$ or $b''_g$ in the case of $v_{\alpha\alpha}$ or $v_{\alpha\beta\gamma\delta}$,
respectively. Thus the variance ratios converge to universal shape-independent constants
in the $kL \to \infty$ limit, but the convergence is logarithmically slow. For example,
in the presence of time-reversal symmetry,
\begin{equation}
\overline{\delta v^2_{\alpha\alpha}} / \overline{\delta v^2_{\alpha\beta}} =6
+ {b'_g-b_g \over \ln kL} +\cdots
=6-{2.15 \over \ln kL} +\cdots
 \label{vaaratio} \,,
\end{equation}
to leading order in $1/\ln kL$. For a disk, this ratio barely reaches $3$ in the $kL$ range of experimental
interest. This is our first indication that beautiful analytic results valid as $kL \to \infty$
may not always have relevance to experiments, even when the dot contains hundreds or thousands of electrons.

\subsection{One-body matrix elements}\label{one-body}

 When an electron is added to the finite dot, charge accumulates on the surface
and its effect can be described by a one-body potential energy ${\cal V}(\vec
r)$. The diagonal matrix elements of ${\cal V}(\vec r)$ are given by
\begin{equation}
\label{valphadef}
v_\alpha \equiv {\cal V}_{\alpha \alpha} =\int_V d
\vec r \; |\psi_\alpha(\vec r)|^2 \, {\cal V}(\vec r) \,.
\end{equation}
Again, we wish to express the variance in terms of the wave
function intensity correlator:
\begin{equation}
\label{v1expr}
\overline{\delta v_\alpha^2} = \int_V \int_V d\vec r \, d\vec r' \; {\cal
V}(\vec r) \,C(\vec r,\vec r') \,{\cal V}(\vec r')  \label{v1expr1}
 \,,
\end{equation}
Because
only one power of $C$ appears in the variance, the
integral is dominated by distant pairs of points $|\vec r-\vec r'| \sim L$, and
scales as $1/kL$:
\begin{equation}
\overline{\delta v_\alpha^2} ={ c_g \over \beta} {\Delta^2 \over kL}
+O\left({\Delta^2 \over (kL)^2}\right) \,,
\label{v1power}
\end{equation}
where $c_g$ is a shape-dependent dimensionless coefficient.
We note that normalization-related subtraction of the correlator, which had only a moderate
effect on the two-body matrix element fluctuations, here reduces the variance
by a full order of magnitude, resulting in a very small dimensionless coefficient: $c_g=0.035$ for a disk, and even smaller
for less symmetric shapes.
This is due the fact that the integrand (\ref{v1expr1}) ceases to be everywhere positive
after subtraction (\ref{corrnorm}), in contrast with the integrand in
(\ref{vabint}). The analytic expression (\ref{v1power}) is in
excellent agreement with exact Monte Carlo simulations (not shown).
Due to the small size of the coefficient $c_g$, the numerical
value of $\overline{\delta v_\alpha^2}$ is smaller
in the physically interesting $kL$ regime than the corresponding result for the
two-body matrix element variance $\overline{\delta v_{\alpha\beta}^2}$, despite the fact that the former is parametrically
larger in a $1/kL$ expansion.

\subsection{Matrix element distributions}
\label{secdistr}

 The central limit theorem implies that all interaction matrix elements in the random wave model must be distributed as Gaussian random
variables as $kL \to \infty$.  This
justifies our focus so far on the variance (and covariance) of these matrix
elements, to the neglect of higher moments.  However, we have seen above that
non-universal finite $kL$ effects are sometimes significant in the
experimentally relevant regime $kL \le 70$, e.g., for the variance
ratio (\ref{vaaratio}).  Thus, we should look explicitly at the shape of matrix element
distributions for finite $kL$.

\begin{figure}[ht] 
{
\includegraphics[height=.3\textheight,angle=270]{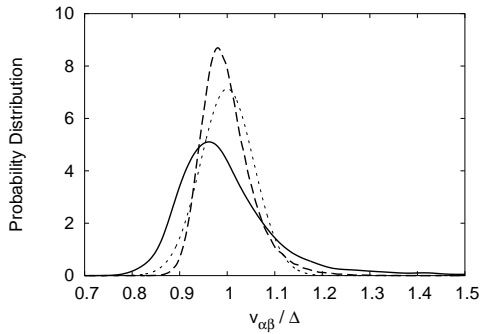}
}

\protect\caption{
The distribution of diagonal interaction matrix elements $v_{\alpha\beta}$ is
shown for real random waves in a disk (dashed curve) and for actual eigenstates
in a modified quarter-stadium billiard geometry (solid curve, see
below) at $kL=70$.  A Gaussian distribution with the same mean
and variance as the random wave distribution is shown as a dotted curve for
comparison.
}
\label{figdistr}
\end{figure}

Data for the diagonal matrix elements $v_{\alpha\beta}$ in
shown in Fig.~\ref{figdistr}.  Results for other
shapes and for other matrix elements (e.g., $v_{\alpha\alpha}$,
$v_{\alpha\beta\gamma\delta}$, or $v_{\alpha}$) are qualitatively similar.  We
note that the numerically obtained interaction matrix element distribution for
random waves (dashed curve) has a long tail on the right side as compared with
a Gaussian distribution of the same mean and variance (dotted curve).  In other
words, there is an excess of anomalously large matrix elements, compensated for
by a reduction in the median to a value slightly below $\Delta$.

Deviations from a Gaussian shape can be quantified by considering higher
moments.
For large $kL$, we may estimate these higher moments in a manner analogous
to our estimate for the variance in Eq.~(\ref{vabint}), but using $n-$point correlators.
After some calculation~\cite{kapalhinprep},
we obtain the skewness
\begin{equation}
\gamma_1 = { \overline{\delta v_{\alpha\beta}^3} \over 
\;\;\;\;\left[\;\overline{\delta v_{\alpha\beta}^2}\;\right]^{3/2} } =
b_{3g} \; c_{3\beta}^2 \left({\beta \over 2} \right)^3
\left({\pi \over 3}\right )^{3/2} (\ln kL)^{-3/2}
\label{gam3}
\end{equation}
and excess kurtosis
\begin{equation}
\gamma_2 = { \overline{\delta v_{\alpha\beta}^4} - 3
 \left[\; \overline{\delta v_{\alpha\beta}^2}\;\right]^2
\over 
\left [\overline{\delta v_{\alpha\beta}^2}\; \right] ^2 }
=b_{4g}\left(c_{4\beta}^2 + \left({2 \over \beta}\right)^4 \right) 
\left({\pi^2 \over 3}\right) (\ln kL)^{-2}\,.
\label{gam4}
\end{equation}
Because the decay is only logarithmic, $\gamma_1$ and
$\gamma_2$ always remain $\ge 1$ for values of $kL$ relevant in the experiments.
The same holds true for other matrix elements and for the
higher moments.  Therefore, matrix element distributions are predicted to be
strongly non-Gaussian, even within the random wave model.

\section{Chaotic Billiards}
\label{truechaos}

\begin{figure}[ht] 
{
\includegraphics[height=.3\textheight,angle=270]{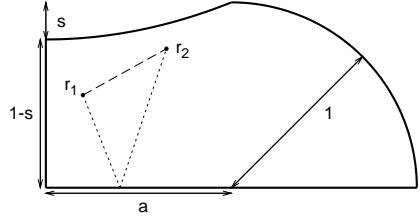}
}

\protect\caption{
A modified quarter-stadium geometry with parameters $a$ and $s$ is used to
illustrate dynamical effects on matrix element fluctuations.  In the Figure, we
set the quarter-circle radius $R=1$.  The random wave contribution to the
correlator, $C^{\rm rw}(\vec r_1,\vec r_2)$ is schematically indicated by a dashed line,
and a typical dynamical contribution by a dotted line.
}
\label{figstadpict}
\end{figure}

  We now investigate how dynamical effects may modify the fluctuations
of interaction matrix elements beyond the (normalized) random wave model.  As an example,
we use a modified
quarter-stadium billiard geometry~\cite{stadium}, where the quarter-circle has
radius $R$ and the straight edge of length $a$ has been replaced by a
parabolic bump to eliminate bouncing-ball modes (Fig.~\ref{figstadpict}).
The system has been verified numerically to be fully
chaotic for the range of parameters used.  Variation of the bump size $s$
allows us to check the sensitivity of the results to details of the billiard
geometry while maintaining the chaotic character of the classical dynamics.
Furthermore, variation of the parameter $a$ allows us to control the degree of
classical chaos.

\begin{figure}[ht] 
{
\includegraphics[height=.3\textheight,angle=270]{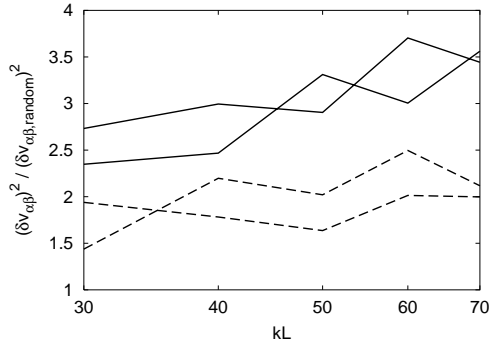}
}

\protect\caption{
The enhancement of the $v_{\alpha\beta}$ variance over the random wave
prediction is shown for sevarl billiards with Neumann
boundary conditions. Solid lines: $a=0.25$, $s=0.1$, $0.2$.  Dashed lines: $a=1.0$, $s=0.1$, $0.2$.
}
\label{figstadvabratio}
\end{figure}

 We first study the variance of the diagonal interaction matrix elements
$v_{\alpha\beta}$.  Typical numerical results are shown in
Fig.~\ref{figstadvabratio}.  We note the large enhancement of the billiard results
over the random wave model.  To understand this discrepancy, we first
compare the exact numerical result for $\overline{\delta v_{\alpha \beta}^2}$ with the
first term on the right hand side of Eq.~(\ref{vabint}), where the intensity
correlator for normalized random waves
 $\tilde C(\vec r, \vec r')$ is replaced by the billiard
correlator $C^{\rm bill}(\vec r,\vec r')$ (calculated numerically for
the appropriate billiard system).  The discrepancy is immediately reduced to a
$\sim 5-10\%$ level, which is comparable to the $O(1/(kL)^{3})$ discrepancy
observed in the random wave model (see inset of Fig.~\ref{figrandvab}).  Thus,
the large enhancement of $v_{\alpha\beta}$ fluctuations over the random wave
prediction  can be
traced directly to an enhancement in the single-particle intensity correlator $C^{\rm
bill}(\vec r,\vec r')$ over $\tilde C(\vec r,\vec r')$.

 Can this dynamical enhancement of the intensity correlator (as compared with a random
wave model) be obtained using a semiclassical approach?  The random wave
correlator $C^{\rm rw}(\vec r,\vec r')$ may be interpreted semiclassically as
arising from straight-line free propagation, as indicated by the
dashed line in Fig.~\ref{figstadpict}.  As discussed by several authors~\cite{srednicki,urbina},
additional contributions to the correlator in a specific
dynamical system can be associated with classical trajectories that bounce off
the boundary $n$ times on their way from $\vec r$ to $\vec r'$, such as the one
indicated by a dotted line in the same figure.  We obtain~\cite{kapalhinprep}
\begin{eqnarray}
\label{vabintsc}
&\!\!\!\!\!\!\!\!\!\!\!\!\!
\overline{\delta v_{\alpha\beta}^2} & \!= \! \Delta^2 V^2 \!\!
\int_V \int_V \! d\vec r \, d\vec r' \,
 (C^{\rm bill}(\vec r,\vec r'))^2 +O\left({\Delta^2 \over (kL)^3}\right) \\
&& \!=\! \Delta^2 {3 \over \pi} \left({2 \over \beta} \right)^2 {(\ln kL
+b_g)+b_{\rm sc}\over (kL)^2} + O\left({\Delta^2 \over (kL)^3}\right) 
\label{bsc}
\end{eqnarray}
where $b_{\rm sc}$ is a classical constant that in practice must be determined
numerically by performing the integral in Eq.~(\ref{vabintsc}).
Random wave and semiclassical contributions to $C^{\rm bill}(\vec
r,\vec r')$ are of the same order except for $|\vec r-\vec r'| \ll L$; it is
these short-distance pairs that result in a logarithmic enhancement of the
random-wave term.

The coefficient $b_{\rm sc}$ may
in practice be quite large, even for generic chaotic systems, such as the
modified stadium billiard. In a diagonal approximation,
$b_{\rm sc} \sim T_{\rm clas}^2$ where $T_{\rm clas}$
is the decay time of classical correlations. Qualitatively, this is consistent
with  Fig.~\ref{figstadvabratio}, as fluctuations are observed to be
consistently larger for the less chaotic $a=0.25$ billiard, as compared with
the $a=1.00$ billiard. We note that both billiards are ``generic", in the
sense that they are not fine-tuned to obtain an anomalously long time scale
$T_{\rm clas}$.  We also note that varying the bump size $s$ has a very
weak effect on the matrix element statistics (as long as $s$ is large enough to
destroy the bouncing-ball modes) and serves instead to provide an estimate of
the statistical fluctuations.

However, a closer look at the data shows that the
numerical results cannot be explained fully by semiclassical arguments,
since (\ref{bsc}) converges to the random wave limit (\ref{bcoeff})
as $kL \to \infty$, while the enhancement factor in Fig.~\ref{figstadvabratio}
{\it grows} with $kL$ in the parameter range of experimental interest.

\begin{figure}[ht] 
{
\includegraphics[height=.3\textheight,angle=270]{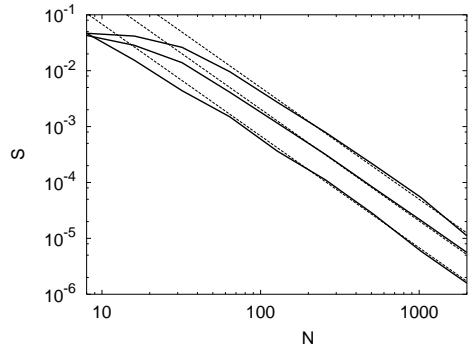}
}

\protect\caption{
The two-body matrix element variance $S$ for a quantum map,
as a function of Hilbert space dimension $N$.  From top to
bottom, the three solid lines represent data for dominant orbit stability
exponent $\lambda_0=0.25$, $0.50$, $1.00$.  The three dashed lines indicate the
asymptotic $1/N^2$ behavior for each case in the semiclassical regime of large
$N$.}
\label{figmap}
\end{figure}

 This anomalous behavior results from a combined effect of two
factors: the large numerical value of $b_{\rm sc}$
for generic dynamical systems, and saturation of the $1/(kL)^2$ behavior at moderate
($< 100$) values of $kL$. As the classical system becomes less unstable and
the correlation time $T_{\rm corr}$ increases, $b_{\rm sc}$ also increases,
leading to greatly enhanced matrix
element variance at very large values of $kL$ (\ref{bsc}).  Because the
variance is bounded above independent of $kL$, the $(kL)^{-2}$ growth in the
variance necessarily breaks down at moderate values of $kL$.  This
saturation sets in at ever larger values of $kL$ as the system
becomes less unstable.  Alternatively,
one may note that the natural expansion parameter for interaction matrix
element fluctuations in a dynamical system is not $1/kL$ but rather the
inverse Thouless conductance $g_T^{-1} \sim T_{\rm corr}/kL$, and
the semiclassical contribution with prefactor $b_{\rm sc}$
in Eq.~(\ref{bsc}) is the leading $O(g_T^{-2})$ effect in such an expansion.
Terms of third and higher order in $g_T^{-1}$, although formally subleading and
not included in a semiclassical calculation, become quantitatively as large as
the leading $O(g_T^{-2})$ term when $g_T$ falls below some characteristic
value.  Furthermore, if one considers chaotic billiards with a long correlation
decay time $T_{\rm corr}$, the importance of formally subleading terms in
the $g_T^{-1}$ expansion will extend to quite large values of $kL$.

 The above assertions are explicitly confirmed for a quantum map model,
which has scaling behavior analogous to
that of a two-dimensional billiard, with the number of states $N=2\pi/\hbar$
playing the role of semiclassical parameter $kL=pL/\hbar$ in the
billiard~\cite{map1,map2}. As in the billiard, a free parameter $\lambda_0$
in the
definition of the map allows for control of the classical correlation decay
time $T_{\rm corr}$.  We see in Fig.~\ref{figmap} that the expected
$N^{-2}$ behavior of the variance is observed at sufficiently large $N$, for
all three families of maps considered.  Furthermore, the prefactor multiplying
$N^{-2}$ in each case agrees with that obtained from a semiclassical
calculation, and as expected this prefactor grows with increasing classical
correlation time $T_{\rm corr}$ (corresponding to a decrease in the
chaoticity of the system).  We also see in Fig.~\ref{figmap} that even for a
``typical" chaotic system (i.e., $T_{\rm corr} \sim 1$), strong deviations
from the $1/N^2$ law appear already below $N \approx 80$.  Such deviations
extend to even larger $N$ for chaotic systems with slower classical correlation
decay.  This suggests that the large-$N$ or large-$kL$ expansion, though
theoretically appealing and asymptotically correct, is problematic in
describing the quantitative behavior of interaction matrix element fluctuations
for real chaotic systems in the physically interesting energy range.

The behavior of $v_{\alpha\alpha}$, $v_{\alpha\beta\gamma\delta}$, and $v_{\alpha}$ in real
dynamical systems may be studied similarly, again using the normalized random wave predictions
as the baseline for comparison.  The enhancement of the variance at large $kL$
is particularly dramatic in the
case of double-diagonal matrix element fluctuations. This is consistent with the reasonable expectation
that dynamical effects lead to particularly strong deviations from random wave
behavior in a modest fraction of the total set of single-particle states, such
as those associated with scarring on unstable periodic
orbits~\cite{hellerscar}. Such deviations lead to a significant tail in the
$v_{\alpha\alpha}$ distribution, but have a minimal effect on the distribution
of off-diagonal matrix elements, since it is unlikely for all four wave
functions $\psi_\alpha$, $\psi_\beta$, $\psi_\gamma$, and $\psi_\delta$ to be
strongly scarred or antiscarred on the same orbit.

We can also go beyond the
variance to investigate higher moments of the matrix element distribution for
actual chaotic systems.  A typical distribution for diagonal two-body
matrix elements $v_{\alpha\beta}$ in a modified quarter-stadium billiard with
$a=0.25$ and $s=0.1$ is shown in Fig.~\ref{figdistr} (solid line).  Since the approach to
Gaussian behavior is already very slow in the case of random waves, it is not
surprising to find even stronger deviations from a Gaussian shape for matrix
elements in real chaotic systems at the same energies.  For example, the excess kurtosis
$\gamma_2$ increases from $8.3$ at $kL=70$ to $20.9$ at $kL=140$, while
dropping from $3.7$ to $3.3$ in the random wave model.  Similar behavior is
obtained for the skewness and for other matrix elements.  Clearly, the distribution tails are very
long, and the assumption of Gaussian matrix element distributions is even less
justified for real chaotic systems than it was in the random wave model.

\section{Beyond the Chaotic Regime}
\label{beyond}

 Finally, we consider fluctuations of matrix elements in systems that
are not fully chaotic.  Here no universal behavior is expected but we shall see
that in such systems the variance can be enhanced much more than in fully
chaotic systems~\cite{ullmo}.  We use the modified quarter-stadium billiard
with $s=0$ or $a<0$.  The choice $s=0$
corresponds to the original Bunimovich stadium, whose quantum fluctuation
properties are dominated by the marginally-stable bouncing-ball modes, while
$a<0$ corresponds to a lemon billiard, which has a classically mixed,
or soft chaotic, phase space. 

In contrast with the $\ln kL/ (kL)^2$ falloff in the $v_{\alpha\beta}$
variance predicted for fully chaotic dynamics by Eq.~(\ref{bsc}), in the case
of regular or mixed dynamics we expect $kL$-independent matrix element
fluctuations of order unity~\cite{kapalhinprep}. Equivalently,
we have very large enhancement, scaling as
$(kL)^2/\ln kL$, of the matrix element variance in mixed dynamical systems,
over the random wave prediction.

\begin{figure}[ht] 
{
\includegraphics[height=.3\textheight,angle=270]{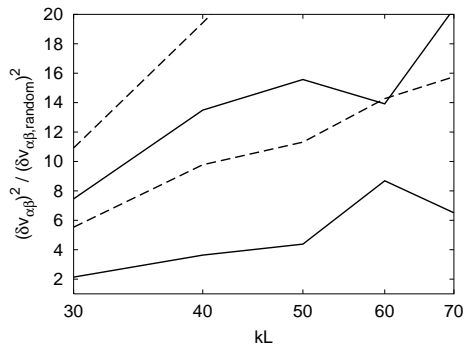}
}

\protect\caption{
Enhancement of $v_{\alpha\beta}$ variance as compared with the random wave
prediction for $a=0.25$, $1.00$ quarter-stadium billiards (solid lines);
$a=-0.25$, $-0.50$ quarter-lemon billiards (dashed lines).  }
\label{figvabmixratio}
\end{figure}

 The diagonal matrix element variance $\overline{\delta v_{\alpha\beta}^2}$ is
computed as a function of $kL$ for two typical mixed phase-space quarter-lemon
billiards and compared with the normalized random wave prediction (dashed lines in Fig.~\ref{figvabmixratio}).
As expected, the enhancement becomes more pronounced at larger $kL$.  Enhancement of an order of magnitude or more over random wave behavior can
easily be obtained for physically interesting values of $kL$.  The most
dramatic enhancement is observed for the $a=-0.25$ quarter-lemon billiard,
which is closer to integrability.

Behavior intermediate between hard chaos and mixed chaotic/regular phase space
is obtained in the presence of families of marginally stable classical
trajectories, such as the ``bouncing ball" orbits of the stadium billiard.
In the quarter stadium billiard ($s=0$ in Fig.~\ref{figstadpict}),
exceptional states associated with such orbits are concentrated in the
rectangular region of the billiard and constitute a fraction $\sim
1/(kL)^{1/2}$ of the total set of states~\cite{baecker}.  When $\alpha$ and
$\beta$ are both bouncing ball states, $\delta v_{\alpha\beta} \sim
\overline{v_{\alpha\beta}} \sim \Delta$, just as would be the case for regular
states concentrated in a finite fraction of the available coordinate space.
These special matrix elements dominate the variance, leading to
\begin{equation}
\overline{\delta v_{\alpha\beta}^2} \sim {\Delta^2 \over kL} \,,
\label{bouncebound}
\end{equation}
Numerical data for quarter-stadium billiards is shown by solid
lines in Fig.~\ref{figvabmixratio}.  The stronger
fluctuations are observed in the less chaotic $a=0.25$ stadium.

Again, analogous results are obtained for the other matrix elements (not shown).

\section{Summary}

 We have studied fluctuations of two-body and one-body matrix elements in
ballistic quantum dots as a function of semiclassical parameter $kL$.
Understanding the quantitative behavior of these fluctuations is important for
the proper analysis of peak spacing statistics in the Coulomb blockade regime. 

 We find that the variance and higher cumulants of two-body and one-body
matrix elements may be expressed in terms of spatial correlations within
single-particle Hartree-Fock wave functions.  For the purpose of computing
the variance of two-body matrix elements $v_{\alpha\alpha}$, $v_{\alpha\beta}$,
and $v_{\alpha\beta\gamma\delta}$, these correlations may be approximated, to
leading order in $kL$ for a chaotic system, by correlations given by
Berry's random wave model.  The calculation results in a variance scaling as
$\ln kL/ (kL)^2$, with universal prefactors depending only on the symmetry
class of the system.  Shape-dependent effects on the variance enter at
$O(1/(kL)^2)$, where the random wave intensity correlator must be corrected
to satisfy individual wave function normalization in finite volume.  In the
normalized random wave model, ratios such as $\overline{\delta
v_{\alpha\alpha}^2}/\overline{\delta v_{\alpha\beta}^2}$ converge only with a
logarithmic rate in the $kL \to \infty$ limit; as a result, the asymptotic
values of such ratios are of little relevance in the regime of experimental
interest.

 The variance of one-body matrix elements $v_{\alpha}$ is affected by
normalization even at leading order, resulting in $O(1/kL)$ scaling in a random
wave model, with a shape-dependent prefactor.  Both two-body and one-body matrix elements
follow, already within the random wave model, a strongly non-Gaussian
distribution, for all physically reasonable values of $kL$.  Thus, higher cumulants
of these matrix elements will be important in peak spacing calculations,
especially in the case of two-body matrix elements where we show that
the approach to a Gaussian distribution is logarithmically slow.

 Dynamical effects, associated with nonrandom short-time behavior in actual
chaotic systems, are formally subleading for two-body matrix
elements, and of the same order as the random wave prediction for one-body
matrix elements. In practice, however, these effects can easily lead to
enhancement by a factor of $3$ or $4$ of the variance in both one-body and
two-body matrix elements, for experimentally relevant values of $kL$ in
reasonable hard chaotic geometries.  The size of these effects scales in
each case as a power of $T_{\rm clas}$, a time scale associated with
approach to ergodicity in the corresponding classical dynamics.  Random wave
behavior is recovered as $T_{\rm clas} \to 0$. In typical geometries,
dynamical effects on matrix element fluctuations cannot be properly computed in
a semiclassical approximation, as higher-order terms are quantitatively of the
same size as the leading-order semiclassical expression in the $kL$ range of experimental
interest.  The approach to semiclassical scaling at very large values of $kL$
as well as saturation of matrix element fluctuations at moderate to small $k$L
are investigated in the context of a quantum map model.

 Systems with a mixed chaotic-regular phase space or with families of
marginally stable classical orbits show even stronger enhancement of matrix
element fluctuations as compared with the random wave model.  The expected
asymptotic scaling with $kL$ of the fluctuations in these cases is discussed,
and is very different from the scaling found in chaotic systems.

 Our calculations strongly indicate that statistics of actual chaotic
single-particle systems, including dynamical effects, are needed to make a
proper comparison between Hartree-Fock-Koopmans theory and experiment.  A
better understanding of correlations in single-particle systems is then
essential to compute observable properties of the interacting many-electron
system.  Furthermore, these correlations need to be understood beyond the naive
leading order semiclassical approximation, to allow comparison with
experiments, which are generally performed at moderate values of the
semiclassical parameter.

\section*{Acknowledgments} We acknowledge useful discussions with Y.~Gefen,
Ph.~Jacquod, and
C.\ H.~Lewenkopf.  This work was supported in part by the U.S. Department of
Energy Grants No.\ DE-FG03-00ER41132 and DE-FG-0291-ER-40608.

\end{document}